\documentclass[a4paper,12pt,showkeys]{revtex4}
\usepackage[utf8]{inputenc}
\usepackage[T1]{fontenc}
\usepackage{lmodern}
\usepackage{epsfig}
\usepackage{array}
\usepackage{amsmath}
\usepackage{epstopdf}
\usepackage{graphicx}
\graphicspath{/}
\begin{document}

\title{Damped oscillation regular structures from the deuteron "effective" electromagnetic form factor data.}
\date{\today}

\author{ A.-Z. Dubni\v ckov\'a$^1$, S. Dubni\v cka$^2$ and P.Weisenpacher$^3$}
\address{$^1$Department of Theoretical Physics, Comenius University, Bratislava, Slovak Republic,\\
$^2$Institute of Physics, Slovak Academy of Sciences, D\'ubravsk\'a cesta 9, SK-84511 Bratislava, Slovak Republic,\\
$^3$Institute of Informatics, Slovak Academy of Sciences, Bratislava, Slovak Republic}
\date{\today}

\begin{abstract}
   The deuteron "D" is the simplest nucleus with the spin S=1, therefore its electromagnetic structure is completely described by three different, the charge
$G_C(t)$, magnetic $G_M(t)$ and quadrupole $G_Q(t)$ form factors, where $t=-q^2$ is the momentum transfer squared of the electrons or the deuterons in the elastic scattering of electrons on deuterons. All three deuteron form factors are theoretically related to the functions $A(t)$, $B(t)$ and $T_{20}(t)$ to be numerically evaluated with errors, whereby $A(t)$ and $B(t)$ in a measurement of the differential cross section of elastic scattering of unpolarized electrons on unpolarized deuterons, and $T_{20}$ in measurements of the elastic scattering of the longitudinally polarized electrons, respectively also on polarized deuteron target. The obtained data are utilized to fix parameters of the deuteron electromagnetic form factors to be constructed in the form of the Unitary and Analytic model. Afterwards these form factors are analytically continued into the time-like region, with the aim to predict artificial behavior of the total cross section $\sigma_{tot}(e^+e^- \to D \bar D)(s)$. By means of the latter artificial data with errors on the deuteron "effective" electromagnetic form factor are produced theoretically. Finally, such data render a possibility to investigate the deuteron damped oscillation regular structures.

\end{abstract}

\maketitle

\section{Introduction.}

   The simplest nucleus deuteron, which now plays in the nuclear physics similar role as the hydrogen atom played in the atomic theory, has not been specified in a
simple way historically. First in 1931 the isotop, approximately two times heavier than the hydrogen, has been identified by Birge and Menzel in \cite{BM}. But no concrete consequences from this experiment have been worked out for its nucleus.

   It was comprehensible, because at the beginning of the thirties, there was no satisfactory theory of the nucleus, as it was thought to be composed of the protons and
electrons, since these were the only known charged particles at that time. The electrons were needed to cancel the positive charge of the protons in the nucleus. This idea has been moreover supported by the fact that the nuclei were seen to emit electrons in their $\beta$-decay.

   While J. Chadwick, investigating several tens in one year's time of other experimental works, came to a conclusion, that the observed heavy and very penetrating
radiation, to have one's rise at the collision of $\alpha$-particles with Be,B and Li, must consist of neutral particles to be called neutrons, first time predicted by Rutherford in 1920 theoretically, Werner Heisenberg \cite{WH} formulated a new idea of a bound state of it with the proton called deuteron (as a stable isotop it was discovered by H.C.Urey et al. \cite{UBM} in 1932) whereby he introduced also the new quantum number, the isospin 1/2 for nucleons, with +1/2 projection for the proton and -1/2 for the neutral particle.  At last the J. Chadwick and M. Goldhaber \cite{ChG} in the first experimental photodisintegration of the deuteron into proton and neutron definitely confirmed a correctness of the idea of Werner Heisenberg on the nucleus deuteron to be a bound state of the proton and neutron with binding energy of 2.19 MeV.

    The electromagnetic (EM) structure of any hadron is completely described by the corresponding EM form factors (FFs). The number of FFs depends on the spin of the
considered hadron. Their behaviors in the timelike region as a function of the c.m. energy squared $s>0$ are in principle measured successfully through the electron-positron annihilation into a hadron-antihadron pair, obtaining the total cross sections $\sigma_{tot}(e^+e^- \to h \bar h)$.

    However, there is an essential difference between the charged pion and the charged and neutral K-mesons with the spin "0", the proton and neutron with the spin
"1/2", for which an abundance data on $\sigma_{tot}(e^+e^- \to h \bar h)$ exists and the deuteron with the spin "1", for which the experimental data on the total cross section $\sigma_{tot}(e^+e^- \to D\bar D)$ are still missing. Nevertheless, in this paper it is demonstrated, how one can solve the latter problem, starting from the
existing data on the elastic scattering of electrons on deuterons in the spacelike region and utilizing well elaborated models of the EM structure of hadrons, based on the analyticity and unitarity of the corresponding EM FFs.

   Almost one decade ago a new phenomenon appeared \cite{BTG} in the elementary particle physics. The so-called "damped oscillation regular structures" (DORS) from
the "effective" proton EM FF data to be obtained from the data on the total cross section $\sigma_{tot}(e^+e^- \to p\bar p)$ have been observed. After some time also new data on the process $e^+e^- \to n \bar n$ with the neutrons have been measured \cite{Ablikim1} in a rather broad region of energies too, and in this case alike DORS have been revealed, however, with just opposite behavior. There are conjectures \cite{T-GP} that the origin of DORS are in the quark gluon structure of the protons and neutrons. Then they have to appear also in the case of the deuteron as the latter can be considered as a six quark state as well. Further we concentrate
to an investigation of the deuteron DORS, despite of the fact that $\sigma_{tot}(e^+e^- \to D\bar D)$ has not been measured experimentally till now.

\section{Electromagnetic structure of deuteron.}
   The deuteron EM structure, taking into account the deuteron spin S=1, is described by three independent EM FFs, the charge $G_C(t)$, the magnetic $G_M(t)$ and the
quadrupole EM FF $G_Q(t)$, as their static values correspond to the charge 1, the magnetic moment $\mu_D=0.8574376\pm0.0000004 \mu_N$ and the quadrupole moment $Q_D=0.2859\pm0.0020 fm^2$ of the deuteron as follows

\begin{eqnarray}\label{norm}
     G_C(0)=1;  G_M(0)=\frac{m_D}{m_p}\mu_D;  G_Q(0)=m_D^2 Q_D,
\end{eqnarray}
where $m_D$ is the deuteron mass and $m_p$ is the proton mass. The relation of the deuteron EM FFs with the total cross section
$\sigma_{tot}(e^+e^- \to D\bar D)(s)$ looks as follows \cite{ADD}

\begin{small}
\begin{eqnarray}\label{totcs}
   \sigma_{tot}(e^+e^- \to D\bar D)(s)=\frac{\pi \alpha^2 \beta_D^3}{3s}\big\{\frac{s}{m^2_D}|G_C(s)+G_M(s)+G_Q(s)|^2+2|G_C(s)+\frac{s}{2m^2_D}G_Q(s)|^2+\\
   +|G_C(s)+\frac{s}{2m^2_D}G_M(s)|^2\big\}\nonumber,
\end{eqnarray}
\end{small}
where $\beta_D(s)=(1-\frac{4 m_D^2}{s})^{1/2}$ is a velocity of out going deuterons and antideuterons in the c.m. system of the reaction $e^+e^- \to D\bar D$.

   The expression (\ref{totcs}) has been obtained by an integration of the differential cross section (78) in the paper \cite{CaGa} of N.Cabbibo and R.Gatto.

   Now we construct the Unitary and Analytic (U$\&$A) models \cite{DD} of the deuteron EM FFs, defined on the four-sheeted Riemann surfaces to be generated by two square
root branch points on the positive real axis, one the lowest possible threshold $s_0$, common for all three deuteron EM FFs and another effective inelastic branch points $s^C_{inl}$, $s^M_{inl}$ and $s^Q_{inl}$, different in three deuteron EM FFs, simulating contributions of all other relevant thresholds given by the unitarity condition of the concrete deuteron EM FF (i.e. the imaginary parts of every deuteron FF represented by the U$\&$A model have to be zero on the first sheet of the Riemann surface from $-\infty$ up to the lowest branch point $s_0$ and different from zero just above it, despite of the fact, that instability of vector mesons in EM FFs is guaranteed by the Breit-Wigner form squared), then the reality condition $F^*(s)=F(s^*)$, responsible for an existence of two complex conjugate vector meson resonance poles exclusively placed on the unphysical sheets of the model, it has to be normalized at s=0 (\ref{norm}) and possess the correct asymptotic behavior in accordance with the QCD predictions \cite{BJL}, \cite{CaGr}. Finally, the numerical values of all parameters with errors of such model could be evaluated in simultaneous comparison of the model with existing data, on $A(t)$, $B(t)$ in the spacelike region, practically measured by the differential cross section of elastic scattering of the unpolarized electrons on unpolarized deuterons,
\begin{eqnarray}
  \frac{d\sigma}{d\Omega}=\frac{\alpha^2 E' cos^2\theta/2}{4 E^3 sin^4\theta/2}[A(t)+B(t) tan^2\theta/2]
\end{eqnarray}
and subsequent application of the Rosenbluth method for the numerical determination of $A(t)$ and $B(t)$ with errors, and $T20(t)$ from the measurement of the polarization tensor $T(t)$ of the recoil deuteron in elastic electron-deuteron scattering with a beam of polarized electrons. The unpolarized real functions of the real variable $A(t)$, $B(t)$ are related to deuteron EM FFs in the spacelike region as follows \cite{GVO}
\begin{eqnarray}\label{A}
  A(t)=G_C^2(t)-\frac{t}{6m_D^2}G_M^2(t)+\frac{t^2}{18m_D^4}G_Q^2(t)
\end{eqnarray}
and
\begin{eqnarray}\label{B}
  B(t)= -\frac{t}{3m_D^2}(1-\frac{t}{4m_D^2})G_M^2(t).
\end{eqnarray}

  $A(t)$ depends on all three deuteron EM FFs, while $B(t)$ depends only on $G_M^2(t)$. So all three deuteron EM FFs in $A(t)$ can not be determined from the
unpolarized cross-section measurement only, therefore a complete separation of all three deuteron EM FFs requires the measurement of at least one tensor polarization, which is commonly taken to be $T_{20}$. Its relation to the deuteron FFs, if we define $R_0=A(t)+B(t)tan^2 \theta/2$ and $\eta=\frac{t}{4m^2_D}$, is \cite{GVO} as follows
\begin{eqnarray}
   T_{20}(t,\theta)=-\frac{1}{\sqrt{2} R_0} \{\frac{8}{3} \eta G_C(t) G_Q(t)+\frac{8}{9}\eta^2G^2_Q(t)+\frac{1}{3} \eta [1+2(1+\eta) tan^2 \theta/2]G^2_M(t)\}.
\end{eqnarray}

   Furthe, nevertheless we are not aimed to calculate all three deuteron FFs from $A(t)$, $B(t)$ and $T_{20}(t,\theta)$ data with errors, and their parameters will be
found directly from the simultaneous fitting of the data on $A(t)$, $B(t)$ and $T_{20}$, by means of the U$\$A$ model of deuteron EM structure as briefly summarized above.

   Determination of the numerical values of FFs parameters in that way enable the model of the deuteron EM structure to continue analytically into the timelike region,
which for $s>s_0$ the deuteron EM FFs are starting to be complex, and they predict the total cross section (\ref{totcs}) behavior as a function of the total c.m. energy squared "s" theoretically.

   The most simple starting point for construction of such U$\&$A model of all three deuteron FFs, like in the case of FFs of other hadrons, is the zero-width
vector-meson-dominance (VMD) model \cite{Sak}, based on the effective Lagrangian of the quantum field theory, in which one assumes that the virtual photon (after having become a quark-antiquark pair) couples to the deuteron through a stable vector meson. The EM FFs of deuteron can then be expressed in terms of the "n" vector-meson masses, the coupling constants ratios between the vector meson "v" and the deuteron $f_{vDD}$, and the coupling strength between the virtual photon and the vector meson $g_{\gamma^*v}=e m_v^2/f_v$, and finally summing over all considered "n" vector meson resonances as follows

\begin{eqnarray}\label{VMD}
   F_D(s)=\sum^n_{v=1} \frac{m_v^2}{m_v^2-s}(f_{vDD}/f_v).
\end{eqnarray}

   The relation (\ref{VMD}), however, does not reflect any above mentioned properties of the deuteron U$\&$A model, besides the creation of "n" vector mesons with
quantum numbers of the photon in the $e^+e^-$ annihilation into the deuteron-antideuteron pairs, and that also only in the primitive form considering them to be stable. Each term in (\ref{VMD}) entails the asymptotic behavior of EM FFs suitable, e.g. for only mesons with a number of building quarks $r_q=2$
\begin{eqnarray}
   F_D(s)_{|s|\to\infty}\sim s^{-1},
\end{eqnarray}
but next it is demonstrated a procedure, how one can generalize the sum in (\ref{VMD}) to behave like
\begin{eqnarray}\label{asymptg}
    F_D(s)_{|s|\to\infty}\sim s^{-m}
\end{eqnarray}
with the number "m"  ($1<m\leq n$) determining the asymptotic behavior of EM FFs of any strongly interacting particle with a number of building quarks $r_q>2$.

   Really, first (\ref{VMD}) is transformed into a rational function with a common denominator and the polynomial of $(n-1)$ degree in the numerator. Then
requiring the first $(m-1)$ coefficients from the highest power "s" in the numerator to be zero, a complicated system of linear homogeneous algebraic equations for coupling constant ratios $a_j=(f_{vDD}/f_v)$ is obtained \cite{ADDPW}. Nevertheless, the utilization of the super-convergent sum rules for $Im F(s)$ \cite{ADDPW} and approximating the FF imaginary part by the $\delta$-function, a more simple system of $(m-1)$ linear homogeneous algebraic equations is derived, which are proved \cite{ADDPW} to be equivalent to the previous one. By their solution, together with the normalization conditions $F_D(0)$ (\ref{norm}), one finds \cite{DDW} general forms of the deuteron EM FFs for $(n>m)$
\begin{small}
\begin{eqnarray}\label{GF1}
   F_D(s)=F_D(0) \frac{\prod^m_{j=1} m^2_j}{\prod^m_{j=1}(m^2_j-s)}+\sum^n_{k=m+1}\{\sum^m_{i=1}\frac{m^2_k}{(m^2_k-s)}\frac{\prod^m_{j\neq k,j=1}m^2_j}{\prod^m_{j\neq k,j=1}(m^2_j-s)}\frac{\prod^m_{j\neq i,j=1}(m^2_j-m^2_k)}{\prod^m_{j\neq i,j=1}(m^2_j-m^2_i)}-\\
   -\frac{\prod^m_{j=1}m^2_j}{\prod^m_{j=1}(m^2_j-s)}\}(f_{kNN}/f_k)\nonumber
\end{eqnarray}
\end{small}
and for $(n=m)$
\begin{small}
\begin{eqnarray}\label{GF2}
   F_D(s)=F_D(0) \frac{\prod^n_{j=1} m^2_j}{\prod^n_{j=1}(m^2_j-s)},
\end{eqnarray}
\end{small}
respectively, for which the required asymptotic behavior (\ref{asymptg}) and the normalization (\ref{norm}) are fulfilled automatically.

   Both forms are suitable for the construction of the U$\&$A model of the deuteron EM structure. However, for the latter one has to know the number "m", determining the
asymptotic behavior of the deuteron EM FFs.

   With this aim, a prediction of the perturbative QCD \cite{BJL}, \cite{CaGr} for the deuteron EM structure function $A(t)$
\begin{eqnarray}
   [A(t)]^{1/2}\sim t^{-5}_{|t \to -\infty}
\end{eqnarray}
is utilized, from which the asymptotic behavior of the deuteron charge EM FF
\begin{eqnarray}\label{asymptC}
   G_C(s)\sim s^{-5}_{|_{|s| \to \infty}}
\end{eqnarray}
is deduced, and the magnetic and quadrupole EM FFs of the deuteron are asymptotically falling like
\begin{eqnarray}\label{asymptMQ}
   G_M(s)=G_Q(s)\sim s^{-6}_{|_{|s| \to \infty}}.
\end{eqnarray}

In this manner the number "m" for the deuteron EM FFs is precisely fixed.

   On the other hand, the behaviors of EM FFs of any hadron, as it is always indicated to some extent by the corresponding total cross section $\sigma_{tot}(e^+e^- \to h
\bar h)$ behavior, are formed by existence of the "n" vector meson resonances with quantum numbers of the photon $J^{PC}= 1^{--}$ and with the isospin
$I=0$ or 1. As the deuteron has no isotopic partner, its isospin is $I_D=0$ and its EM FFs can be saturated in (\ref{VMD}) exclusively with 6 existing isoscalar resonances, $\omega(782), \phi(1020), \omega'(1420), \omega{''}(1650), \phi'(1680), \phi{''}(2170)$, with all their characteristics as given in the Review of Particle Physics \cite{PDG}.

   Then, by means of the general form (\ref{GF1}), for $G_C(s)$ with $n=6>m$, as its correct asymptotic behavior is (\ref{asymptC}), one can write down
the following VMD expression
\begin{footnotesize}
\begin{eqnarray}\label{G_C}
  G_C(s)=[1-(f^c_{\phi''D\bar D}/f_{\phi''})]\frac{m^2_\omega}{m^2_\omega-s}\frac{m^2_\phi}{m^2_\phi-s}\frac{m^2_{\omega'}}{m^2_{\omega'}-s}\nonumber
  \frac{m^2_{\omega''}}{m^2_{\omega''}-s}\frac{m^2_{\phi'}}{m^2_{\phi'}-s}+\frac{m^2_{\phi''}}{m^2_{\phi''}-s}\times\\\nonumber
  \times\{\frac{m^2_\phi}{m^2_\phi-s}\frac{m^2_{\omega'}}{m^2_{\omega'}-s}\frac{m^2_{\omega''}}{m^2_{\omega''}-s}\frac{m^2_{\phi'}}{m^2_{\phi'}-s}\frac{(m^2_\phi-m^2_{\phi''})}
  {(m^2_\phi-m^2_\omega)}\frac{(m^2_{\omega'}-m^2_{\phi''})}{(m^2_{\omega'}-m^2_\omega)}\frac{(m^2_{\omega''}-m^2_{\phi''})}{(m^2_{\omega''}-m^2_\omega)}
  \frac{(m^2_{\phi'}-m^2_{\phi''})}{(m^2_{\phi'}-m^2_\omega)}+\\\nonumber
  +\frac{m^2_\omega}{m^2_\omega-s}\frac{m^2_{\omega'}}{m^2_{\omega'}-s}\frac{m^2_{\omega''}}{m^2_{\omega''}-s}\frac{m^2_{\phi'}}{m^2_{\phi'}-s}\frac{(m^2_\omega-m^2_{\phi''})}
  {(m^2_\omega-m^2_\phi)}\frac{(m^2_{\omega'}-m^2_{\phi''})}{(m^2_{\omega'}-m^2_\phi)}\frac{(m^2_{\omega''}-m^2_{\phi''})}{(m^2_{\omega''}-m^2_\phi)}
  \frac{(m^2_{\phi'}-m^2_{\phi''})}{(m^2_{\phi'}-m^2_\phi)}+\\
  +\frac{m^2_\omega}{m^2_\omega-s}\frac{m^2_\phi}{m^2_\phi-s}\frac{m^2_{\omega''}}{m^2_{\omega''}-s}\frac{m^2_{\phi'}}{m^2_{\phi'}-s}\frac{(m^2_\omega-m^2_{\phi''})}
  {(m^2_\omega-m^2_{\omega'})}\frac{(m^2_\phi-m^2_{\phi''})}{(m^2_\phi-m^2_{\omega'})}\frac{(m^2_{\omega''}-m^2_{\phi''})}{(m^2_{\omega''}-m^2_{\omega'})}
  \frac{(m^2_{\phi'}-m^2_{\phi''})}{(m^2_{\phi'}-m^2_{\omega'})}+\\\nonumber
  +\frac{m^2_\omega}{m^2_\omega-s}\frac{m^2_\phi}{m^2_\phi-s}\frac{m^2_{\omega'}}{m^2_{\omega'}-s}\frac{m^2_{\phi'}}{m^2_{\phi'}-s}\frac{(m^2_\omega-m^2_{\phi''})}
  {(m^2_\omega-m^2_{\omega''})}\frac{(m^2_\phi-m^2_{\phi''})}{(m^2_\phi-m^2_{\omega''})}\frac{(m^2_{\omega'}-m^2_{\phi''})}{(m^2_{\omega'}-m^2_{\omega''})}
  \frac{(m^2_{\phi'}-m^2_{\phi''})}{(m^2_{\phi'}-m^2_{\omega''})}+\\\nonumber
  +\frac{m^2_\omega}{m^2_\omega-s}\frac{m^2_\phi}{m^2_\phi-s}\frac{m^2_{\omega'}}{m^2_{\omega'}-s}\frac{m^2_{\omega''}}{m^2_{\omega''}-s}\frac{(m^2_\omega-m^2_{\phi''})}
  {(m^2_\omega-m^2_{\phi'})}\frac{(m^2_\phi-m^2_{\phi''})}{(m^2_\phi-m^2_{\phi'})}\frac{(m^2_{\omega'}-m^2_{\phi''})}{(m^2_{\omega'}-m^2_{\phi'})}
  \frac{(m^2_{\omega''}-m^2_{\phi''})}{(m^2_{\omega''}-m^2_{\phi'})}\}(f^c_{\phi''D\bar D}/f_{\phi''}),\nonumber
\end{eqnarray}
\end{footnotesize}

and for $G_M(s)$ and $G_Q(s)$ the general form (\ref{GF2}), as for them the relation n=m=6 is valid, gives
\begin{footnotesize}
\begin{eqnarray}\label{G_M}
   G_M(s)=\frac{m_D}{m_p}\mu_D \frac{m^2_\omega}{m^2_\omega-s}\frac{m^2_\phi}{m^2_\phi-s}\frac{m^2_{\omega'}}{m^2_{\omega'}-s}
  \frac{m^2_{\omega''}}{m^2_{\omega''}-s}\frac{m^2_{\phi'}}{m^2_{\phi'}-s}\frac{m^2_{\phi''}}{m^2_{\phi''}-s},
\end{eqnarray}
\end{footnotesize}
and
\begin{footnotesize}
\begin{eqnarray}\label{G_Q}
   G_Q(s)=m_D^2Q \frac{m^2_\omega}{m^2_\omega-s}\frac{m^2_\phi}{m^2_\phi-s}\frac{m^2_{\omega'}}{m^2_{\omega'}-s}
  \frac{m^2_{\omega''}}{m^2_{\omega''}-s}\frac{m^2_{\phi'}}{m^2_{\phi'}-s}\frac{m^2_{\phi''}}{m^2_{\phi''}-s}.
\end{eqnarray}
\end{footnotesize}

   In the next step all VMD terms $m^2_r/(m^2_r-s)$ of the previous three expressions will be related to the function

\begin{small}
\begin{eqnarray}\label{confmap}
 V(s)=i\frac{\sqrt{(\frac{s_{inl}-s_0}{s_0})^{1/2}+(\frac{s-s_0}{s_0})^{1/2}}-
   \sqrt{(\frac{s_{inl}-s_0}{s_0})^{1/2}-(\frac{s-s_0}{s_0})^{1/2}}}
   {\sqrt{(\frac{s_{inl}-s_0}{s_0})^{1/2}+(\frac{s-s_0}{s_0})^{1/2}}+
   \sqrt{(\frac{s_{inl}-s_0}{s_0})^{1/2}-(\frac{s-s_0}{s_0})^{1/2}}},
\end{eqnarray}
\end{small}
which is mapping (see FIG.1) the corresponding four-sheeted Riemann surfaces (see FIG.1 left) into one plane (see FIG.1 right) with unit disc, whereby the first physical sheet of the Riemann surface is completely mapped into left half disc, the second sheet is mapped into right half disc, the third sheet into left half plane besides the left half disc and finally the fourth sheet is mapped into right half plane besides the right half disc.

\begin{figure}
    \includegraphics[width=0.25\textwidth]{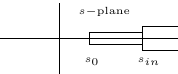}\hspace{0.3cm}
    \includegraphics[width=0.25\textwidth]{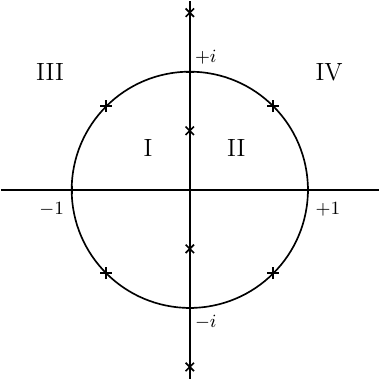}\\
\caption{Complex s-plane with square root braqnch points $s_0$ and $s_{inl}$ left and conformally mapped four sheeted Riemnn surface into one plane with unit
disc right).\label{FIG.1}}
\end{figure}

   Practically it is done by means of the inverse function to (\ref{confmap})
\begin{eqnarray}\label{invrel}
    s=s_0-\frac{4(s^{(C)}_{inl}-s_0)}{(1/V-V)^2};
    s=s_0-\frac{4(s^{(M)}_{inl}-s_0)}{(1/U-U)^2};
    s=s_0-\frac{4(s^{(Q)}_{inl}-s_0)}{(1/W-W)^2},
\end{eqnarray}
which are substituted into the VMD terms in the $G_C(s), G_M(s), G_Q(s)$, by using also the relations
\begin{eqnarray}
    m^2_r=s_0-\frac{4(s^{(C)}_{inl}-s_0)}{(1/V_{r0}-V_{r0})^2};
    m^2_r=s_0-\frac{4(s^{(M)}_{inl}-s_0)}{(1/U_{r0}-U_{r0})^2};
    m^2_r=s_0-\frac{4(s^{(Q)}_{inl}-s_0)}{(1/W_{r0}-W_{r0})^2};
\end{eqnarray}
and
\begin{eqnarray}
    0=s_0-\frac{4(s^{(C)}_{inl}-s_0)}{(1/V_N-V_N)^2};
    0=s_0-\frac{4(s^{(M)}_{inl}-s_0)}{(1/U_N-U_N)^2};
    0=s_0-\frac{4(s^{(Q)}_{inl}-s_0)}{(1/W_N-W_N)^2};
\end{eqnarray}
with $V_N=V(0), U_N=U(0), W_N=W(0)$, following from (\ref{invrel}) automatically.

    Then for any VMD term of (\ref{G_C})-(\ref{G_Q}), one gets the factorized expressions (e.g. taken concretely from $G_Q$, however the obtained results are
valid for all other VMD terms in (\ref{G_C})-(\ref{G_Q}))
\begin{eqnarray}\label{4resp}
    \frac{m^2_r-0}{m^2_r-s}=\big(\frac{1-W^2}{1-W^2_N}\big)^2\frac{(W_N-W_{r0})(W_N+W_{r0})(W_N-1/W_{r0})(W_N+1/W_{r0})}{(W-W_{r0})(W+W_{r0})(W-1/W_{r0})(W+1/W_{r0})},
\end{eqnarray}
into the asymptotic terms $\big(\frac{1-V^2}{1-V^2_N}\big)^2$, $\big(\frac{1-U^2}{1-U^2_N}\big)^2$, $\big(\frac{1-W^2}{1-W^2_N}\big)^2$ and into the so-called finite-energy terms (as for $|s|\to \infty$ they turn out into the real constants), from which one deduces a creation of four independent poles, whereby further their arrangement depends, if the real part of the Breit-Wigner form squared $(m_r-i\Gamma_r/2)^2$ is
\begin{eqnarray}\label{zwdth1}
   {(m^2_r-\Gamma^2_r/4)}<s_{inl} \Rightarrow V_{r0}=-V^*_{r0}, U_{r0}=-U^*_{r0}, W_{r0}=-W^*_{r0}
\end{eqnarray}
then
\begin{eqnarray}\label{pimagax}
    \frac{m^2_r-0}{m^2_r-s}=\big(\frac{1-W^2}{1-W^2_N}\big)^2\frac{(W_N-W_{r0})(W_N-W^*_{r0})(W_N-1/W_{r0})(W_N-1/W^*_{r0})}{(W-W_{r0})(W-W^*_{r0})(W-1/W_{r0})(W-1/W^*_{r0})},
\end{eqnarray}
is producing four poles on the imaginary axis (first two inside of the unit circle and last two outside of the unit circle),
or, if the real part of the Breit-Wigner form squared $(m_r-i\Gamma_r/2)^2$ is
\begin{eqnarray}\label{zwdth2}
   {(m^2_r-\Gamma^2_r/4)}>s_{inl} \Rightarrow V_{r0}=1/V^*_{r0}, U_{r0}=1/U^*_{r0}, W_{r0}=1/W^*_{r0}
\end{eqnarray}
then
\begin{eqnarray}\label{punitcir}
    \frac{m^2_r-0}{m^2_r-s}=\big(\frac{1-W^2}{1-W^2_N}\big)^2\frac{(W_N-W_{r0})(W_N-W^*_{r0})(W_N+W_{r0})(W_N+W^*_{r0})}{(W-W_{r0})(W-W^*_{r0})(W+W_{r0})(W+W^*_{r0})}
\end{eqnarray}
is producing four poles to be symmetrically placed on the unit circle.
    The pure mass terms in (\ref{G_C}) are transformed similarly, however every of them first is decomposed into difference of two terms as follows:
    \begin{eqnarray}
    \frac{m^2_2-m^2_6}{m^2_2-m^2_1}= \frac{m^2_2-0}{m^2_2-m^2_1}-\frac{m^2_6-0}{m^2_2-m^2_1}
    \end{eqnarray}
etc.

    Lastly, introducing the non-zero width of the resonance by the Breit-Wigner form $m^2_r \to (m_r-i\Gamma_r/4)^2$, i.e. simply one has to rid of "0" in sub-indices of
(\ref{pimagax}), the relations
\begin{eqnarray}\label{pimagaxdef}
    \frac{m^2_r-0}{m^2_r-s}=\big(\frac{1-W^2}{1-W^2_N}\big)^2\frac{(W_N-W_{r})(W_N-W^*_{r})(W_N-1/W_{r})(W_N-1/W^*_{r})}{(W-W_{r})(W-W^*_{r})(W-1/W_{r})(W-1/W^*_{r})},
\end{eqnarray}
are obtained in which all four poles from the imaginary axis are shifted to the unphysical sheets, those inside of the unit disc to the II sheet and those outside of the unit disc to he IV sheet.
    Similarly it is happened with (\ref{punitcir})
\begin{eqnarray}\label{punitcirdef}
    \frac{m^2_r-0}{m^2_r-s}=\big(\frac{1-W^2}{1-W^2_N}\big)^2\frac{(W_N-W_{r})(W_N-W^*_{r})(W_N+W_{r})(W_N+W^*_{r})}{(W-W_{r})(W-W^*_{r})(W+W_{r})(W+W^*_{r})}
\end{eqnarray}
in which all poles are shifted out of the unit disc to the third and forth sheets.

    The relations (\ref{zwdth1}) and (\ref{zwdth2}) are simply clarified by means of the transformation (\ref{confmap}). Really, from the latter one obtains
\begin{small}
\begin{eqnarray}\label{confmapr1}
 V(m^2_r)=U(m^2_r)=W(m^2_r)=i\frac{\sqrt{(\frac{s_{inl}-s_0}{s_0})^{1/2}+(\frac{m^2_r-s_0}{s_0})^{1/2}}-
   \sqrt{(\frac{s_{inl}-s_0}{s_0})^{1/2}-(\frac{m^2_r-s_0}{s_0})^{1/2}}}
   {\sqrt{(\frac{s_{inl}-s_0}{s_0})^{1/2}+(\frac{m^2_r-s_0}{s_0})^{1/2}}+
   \sqrt{(\frac{s_{inl}-s_0}{s_0})^{1/2}-(\frac{m^2_r-s_0}{s_0})^{1/2}}},
\end{eqnarray}
\end{small}
and if $s_0<m^2_r<s_{inl}$ then one can see immediately from $(\ref{confmapr1})$ that $V^*(m^2_r)=-V(m^2_r)$, $U^*(m^2_r)=-U(m^2_r)$ and $W^*(m^2_r)=-W(m^2_r)$. However, if $m^2_r>s_{inl}$, then from $(\ref{confmapr1})$ it follows
\begin{small}
\begin{eqnarray}\label{confmapr2}
V(m^2_r)=U(m^2_r)= W(m^2_r)=i\frac{\sqrt{(\frac{s_{inl}-s_0}{s_0})^{1/2}+(\frac{m^2_r-s_0}{s_0})^{1/2}}-
   i\sqrt{(\frac{{m^2_r}-s_0}{s_0})^{1/2}-(\frac{s_{inl}-s_0}{s_0})^{1/2}}}
   {\sqrt{(\frac{s_{inl}-s_0}{s_0})^{1/2}+(\frac{m^2_r-s_0}{s_0})^{1/2}}+
   i\sqrt{(\frac{m^2_r-s_0}{s_0})^{1/2}-(\frac{s_{inl}-s_0}{s_0})^{1/2}}},
\end{eqnarray}
\end{small}
and by a multiplication of the latter with a complex conjugate relation to it, one derives the second relations $V(m^2_r) V^*(m^2_r)=1$, $U(m^2_r) U^*(m^2_r)=1$, $W(m^2_r) W^*(m^2_r)=1$ for an arrangement of (\ref{4resp}) leading to (\ref{punitcirdef}).

    The last point before of an application of the Unitary and Analytic model of the deuteron EM structure is the value of the lowest branch point "$s_0$". Since
the deuteron is isoscalar, the natural value of the lowest branch point on the real axis could be $3m^2_\pi$ threshold. However, the inner structure of the deuteron
provides such triangle Feynman diagram \cite{ADD},\cite{DD} that its dual diagram gives the lowest branch point "$s_0$"value
\begin{eqnarray}
    s_0=4m^2_p-\frac{(m^2_D-m^2_p-m^2_n)^2}{m^2_n}=1.7298 m^2_\pi,
\end{eqnarray}
which is called the anomalous threshold of the deuteron EM FFs to be lower than $3m^2_\pi$ threshold.
\section{ANALYSIS OF THE DATA ON DEUTERON STRUCTURE FUNCTIONS AND DEUTERON "DAMPED" OSCILLATION REGULAR STRUCTURES}

    From an existing literature \cite{BuchYe}-\cite{Abbot2} we have collected 140 experimental points on $A(t)$, 41 on $B(t)$ and 25 on $T_{20}$, as graphically
presented in FIG. 2 and FIG. 3 left, respectively.

\begin{figure}[bth]
    \includegraphics[width=0.4\textwidth]{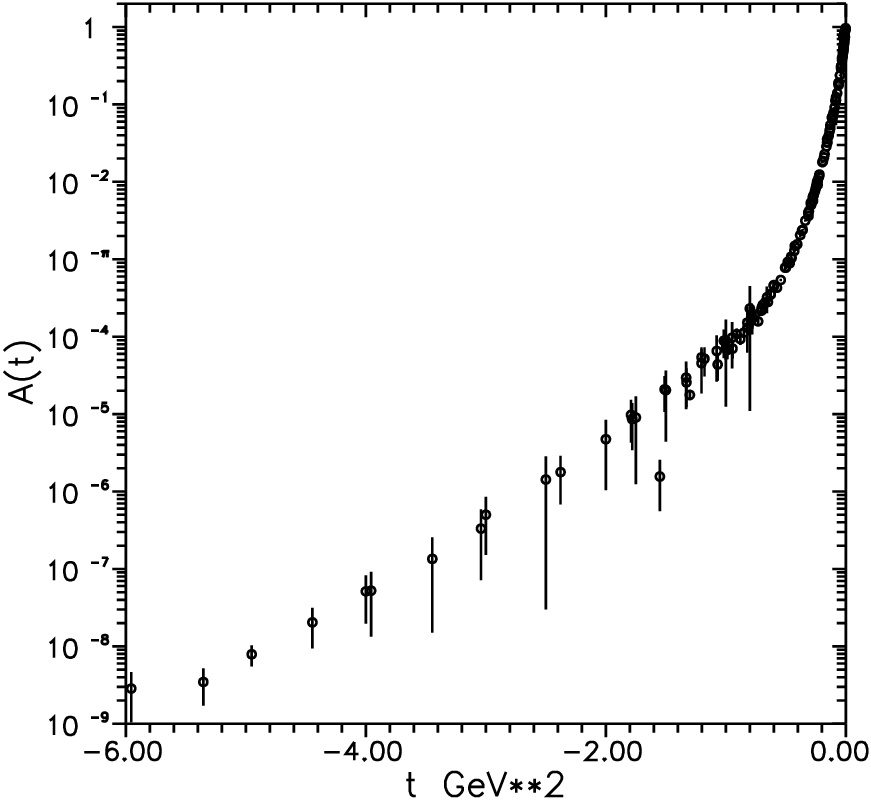}\hspace{0.3cm}
    \includegraphics[width=0.4\textwidth]{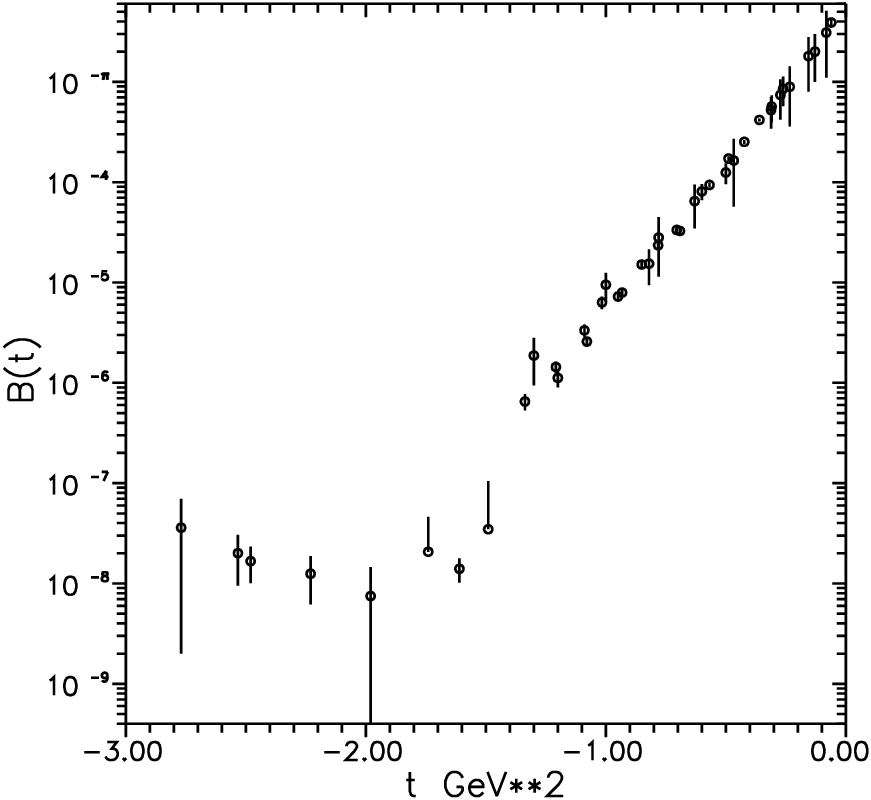}\\
\caption{Existing experimental data on A(t) left and B(t) right in space-like region.\label{fig:2}}
\end{figure}

\begin{figure}[bth]
  \includegraphics[width=0.4\textwidth]{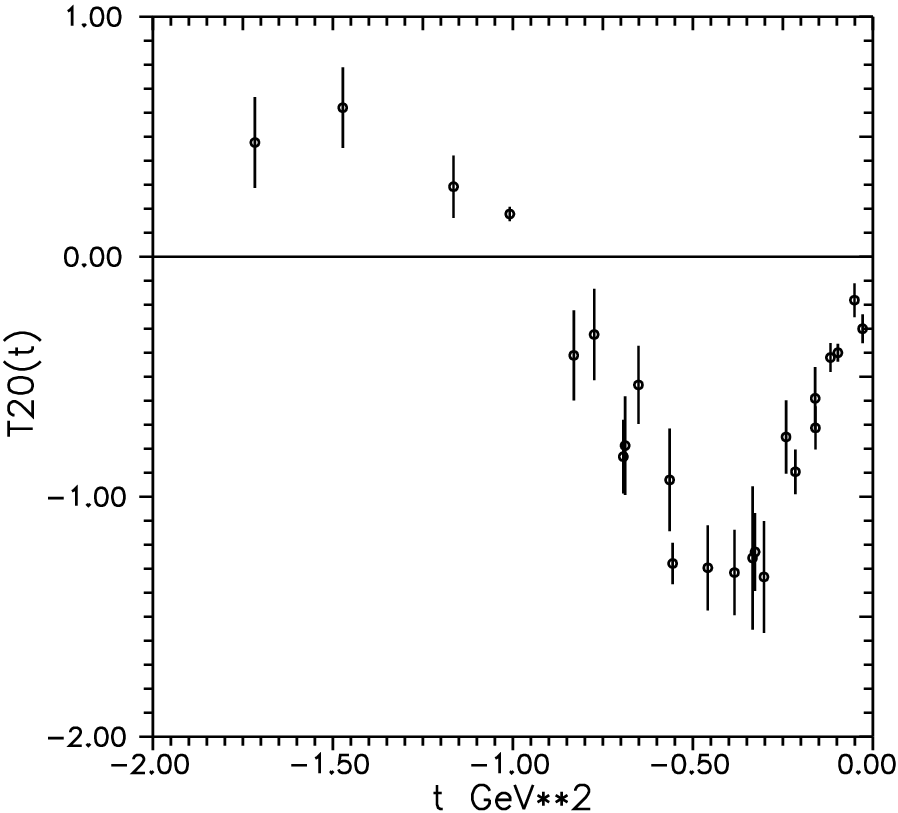}\hspace{0.3cm}
  \includegraphics[width=0.4\textwidth]{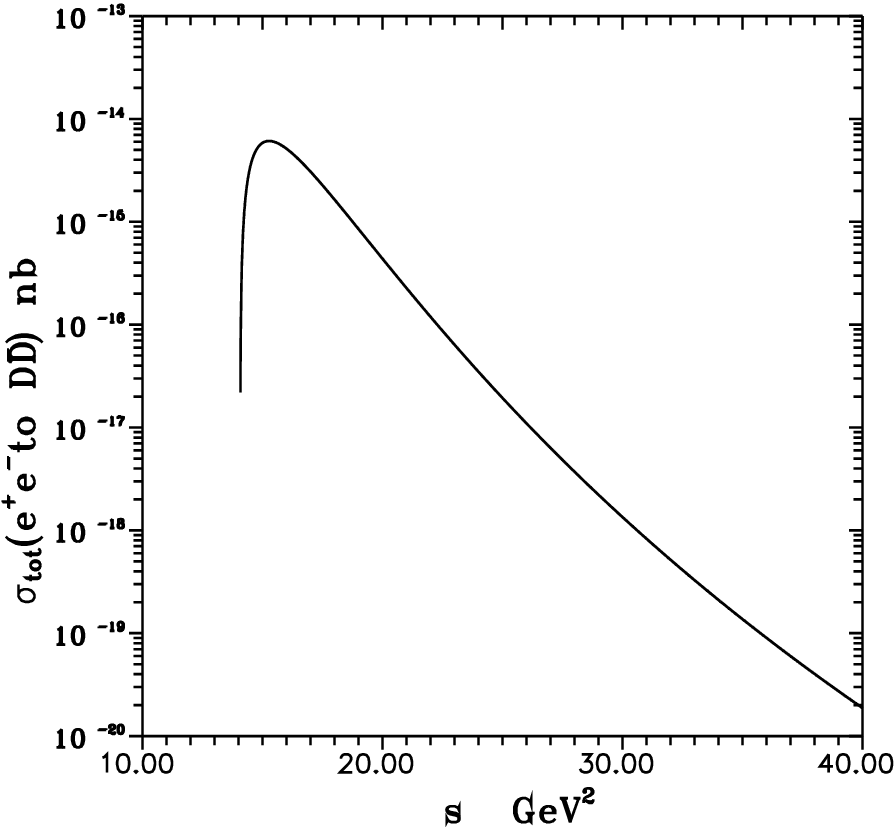}\\
\caption{The data on $T_{20}(t)$ in the space-like region left and predicted behavior of $\sigma_{tot}(e^+e^- \to D\bar D)(s)$ right.}\label{FIG. 3:}
\end{figure}

   The best possible description of the latter data is achieved with $\chi^2/ndf=3.58$ (indicating some inconsistency among the $A(t)$, $B(t)$ and $T_{20}$ data)
and the numerical values of the fitted parameters $s^C_{inl}=4.5857\pm0.0574$, $s^M_{inl}=8.0551\pm0.1092$, $s^Q_{inl}=6.3023\pm0.1190$ and
$(f_{\phi''DD}/f_{\phi''})=0.00443\pm0.00004$, by means of which all three deuteron FFs are specified. Then by their analytic continuation into the timelike region the behavior of the total cross section $\sigma_{tot}(e^+e^- \to D\bar D)$ (\ref{totcs}) could be predicted theoretically, as presented by the curve in FIG. 3 right.

   By means of such curve in FIG. 3 right and the expression
\begin{eqnarray}\label{efff}
     G^{eff}_D(s)=\sqrt{\frac{\sigma_{tot}(e^+e^- \to D\bar D)}{\frac{\pi \alpha^2 \beta_D^3}{s}\big\{\frac{3s}{m^2_D}+(1+\frac{s}{2m^2_D})^2\big\}}}
\end{eqnarray}
 the behavior of the "effective" EM FF of the deuteron is predicted in the form of the line in FIG. 4, in despite of the fact that there exists no one experimental point on the $\sigma_{tot}(e^+e^- \to D\bar D)$ up to now.
\begin{figure}[bth]
  \includegraphics[width=.40\textwidth]{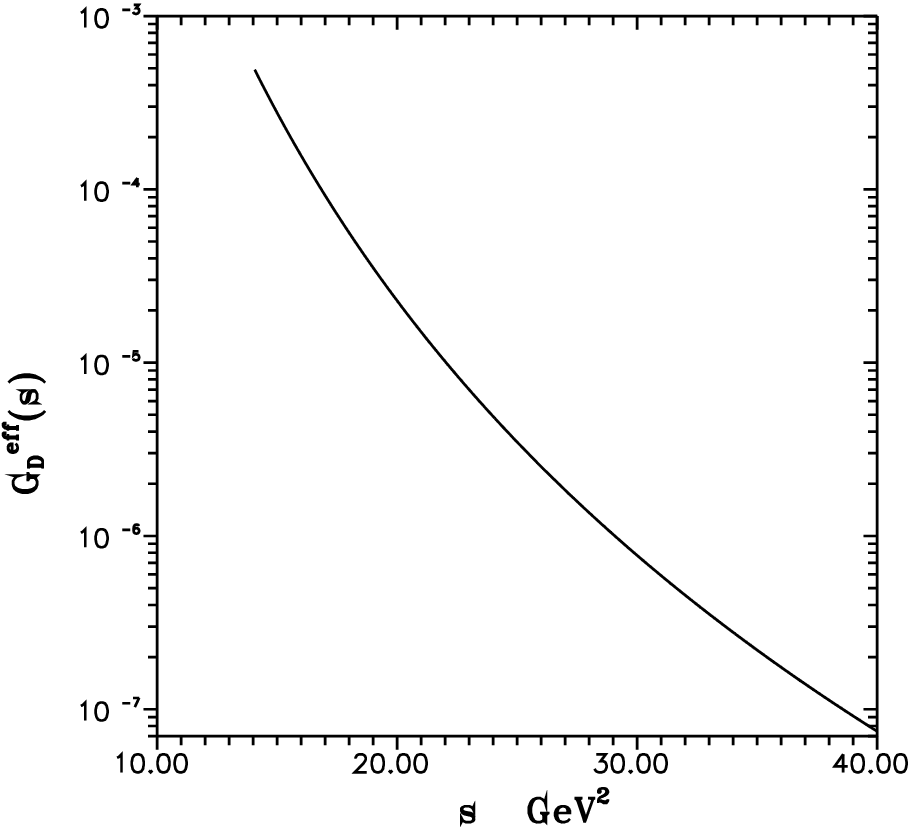}\\
  \caption{Predicted behavior of the deuteron "effective" EM FF $G^{eff}_D(s)$ by means of the expression (\ref{efff}) and $\sigma_{tot}(e^+e^- \to D\bar D)$ presented in FIG. 3 right.}\label{FIG. 4:}
\end{figure}

   Next, an application of the predicted behavior of $G^{eff}_D(s)$ form factor in the form of the line in FIG. 4 is twofold. First, it is applied to a creation of
artificial deuteron "effective" EM FF data with errors in TABLE I , utilizing the proton effective EM FF data with errors presented in FIG. 5, and then also for demonstration that a description of these artificial data by means of the $G^{eff}_D(s)$ line in FIG. 4 is better than their description by the three-parametric function of Tomasi-Gustafsson-Rekalo \cite{TGR} creating DORS.
\begin{figure}[bth]
  \includegraphics[width=.40\textwidth]{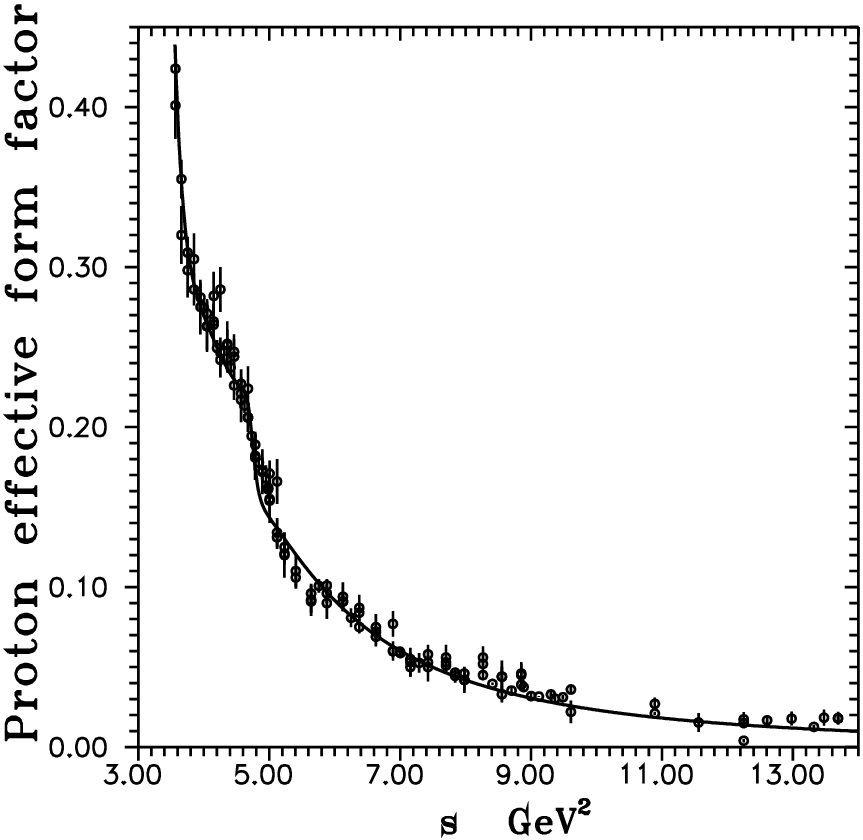}\\
  \caption{By means of the nucleon U$\&$A model predicted curve of the "effective" proton EM FF and its comparison with existing experimental data with errors.}\label{FIG. 5:}
\end{figure}

\begin{figure}[bth]
  \includegraphics[width=.40\textwidth]{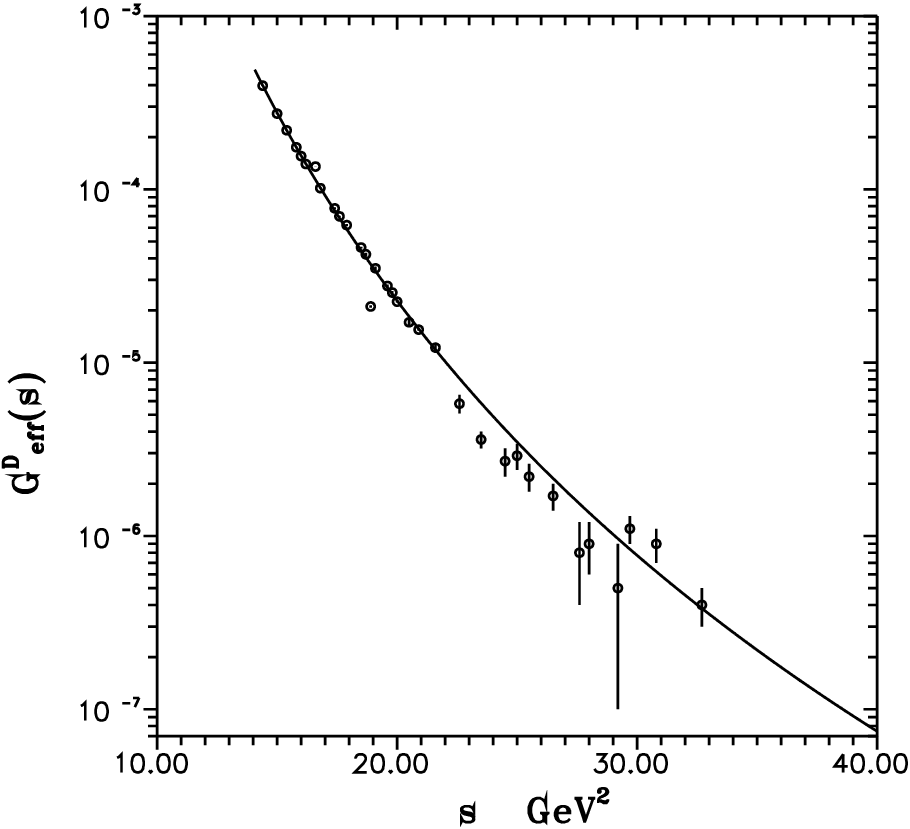}\\
  \caption{Artificial data on the "effective" deuteron EM FF scattered around the theoretically predicted curve by the U$\&$A model of the deuteron EM FFs.}\label{FIG. 6:}
\end{figure}

\begin{table}
\caption{The artificial data on $G^{eff}_D(s)$ with errors from the "effective" proton EM FF in FIG. 5.} \label{TABLE:I}
  s [GeV$^{2}$]   $G^{eff}_D\pm\Delta G^{eff}_D$ \quad s [GeV$^{2}$]   $G^{eff}_D\pm\Delta G^{eff}_D$ \quad s [GeV$^{2}$]   $G^{eff}_D\pm\Delta G^{eff}_D$\\

    $14.36$    $.0004608\pm 0.0000080$    $19.13$    $.0000253\pm 0.0000040$   $23.94$    $.0000035\pm 0.0000030$ \\
    $14.69$    $.0003723\pm 0.0000080$    $19.89$    $.0000213\pm 0.0000060$   $24.80$    $.0000151\pm 0.0000060$ \\
    $14.85$    $.0003399\pm 0.0000080$    $20.67$    $.0000119\pm 0.0000040$   $25.23$    $.0000055\pm 0.0000008$ \\
    $15.02$    $.0002824\pm 0.0000070$    $20.97$    $.0000122\pm 0.0000013$   $25.67$    $.0000091\pm 0.0000060$ \\
    $15.36$    $.0002324\pm 0.0000090$    $21.01$    $.0000108\pm 0.0000014$   $25.11$    $.0000018\pm 0.0000017$ \\
    $15.70$    $.0001845\pm 0.0000070$    $21.47$    $.0000067\pm 0.0000060$   $26.55$    $.0000162\pm 0.0000060$ \\
    $16.92$    $.0000877\pm 0.0000050$    $21.87$    $.0000064\pm 0.0000061$   $26.66$    $.0000013\pm 0.0000010$ \\
    $17.64$    $.0000677\pm 0.0000060$    $22.28$    $.0000212\pm 0.0000060$   $27.00$    $.0000018\pm 0.0000010$ \\
    $18.38$    $.0000457\pm 0.0000060$    $23.10$    $.0000105\pm 0.0000060$   $27.36$    $.0000013\pm 0.0000003$ \\
    $18.75$    $.0000374\pm 0.0000059$    $23.52$    $.0000053\pm 0.0000030$   $32.67$    $.0000074\pm 0.0000040$ \\

\end{table}

   The creation of the artificial data on the "effective" deuteron EM FF from the "effective" proton EM FF data with errors is carried out as follows.

   First, deviations of the proton "effective" EM FF data with errors of the order $10^{-1}$ in FIG. 5 are calculated by a subtraction of the curve in this figure from
the central values of those data at the corresponding c.m. energy squared "s" from the interval 4.7852-10.8900. With the aim of a covering the energetic region of the deuteron in FIG. 4, every value of the c.m. energy squared "s" is multiplied by a factor 3.

   Second, as the curve of $G^{eff}_D(s)$ in FIG. 4 is in averaged of the order $10^{-3}$ - $10^{-6}$, in order to achieve artificial data of the same order, the
obtained deviations of the proton "effective" EM FF data with errors are multiplied by a factor $10^{-3}$. Only such resultant data with errors of the proton "effective" EM FF data are added to the theoretically predicted curve of the deuteron "effective" EM FF in FIG. 4. As a result the 30 artificial data on the "effective" deuteron EM FF is obtained in TABLE I, which are scattered around the theoretically predicted curve of $G^{eff}_D(s)$ as presented in FIG. 6.

   Now in order to investigate the deuteron damped oscillation regular structures (DDORS), first we describe the data in TABLE I optimally by means of the three
parametric function of Tomasi-Gustafsson-Rekalo formula \cite{TGR}

\begin{eqnarray}\label{effFF}
   G^{eff}_D(s)=\frac{A(1)}{(1+\frac{s}{A(2)})(1-\frac{s}{A(3)})^2},
\end{eqnarray}
with the values of the parameters A(1)=0.00308, A(2)=0.06147 $GeV^2$, A(3)=12.24314 $GeV^2$ and the $\chi^2/ndf$=22.8 . The obtained result is presented in FIG. 7. Afterwards this curve is subtracted from the data in TABLE I with a result drawn in FIG. 8, in which one clearly see an existence of the deuteron damped oscillation regular structures.
\begin{figure}[bth]
  \includegraphics[width=.30\textwidth]{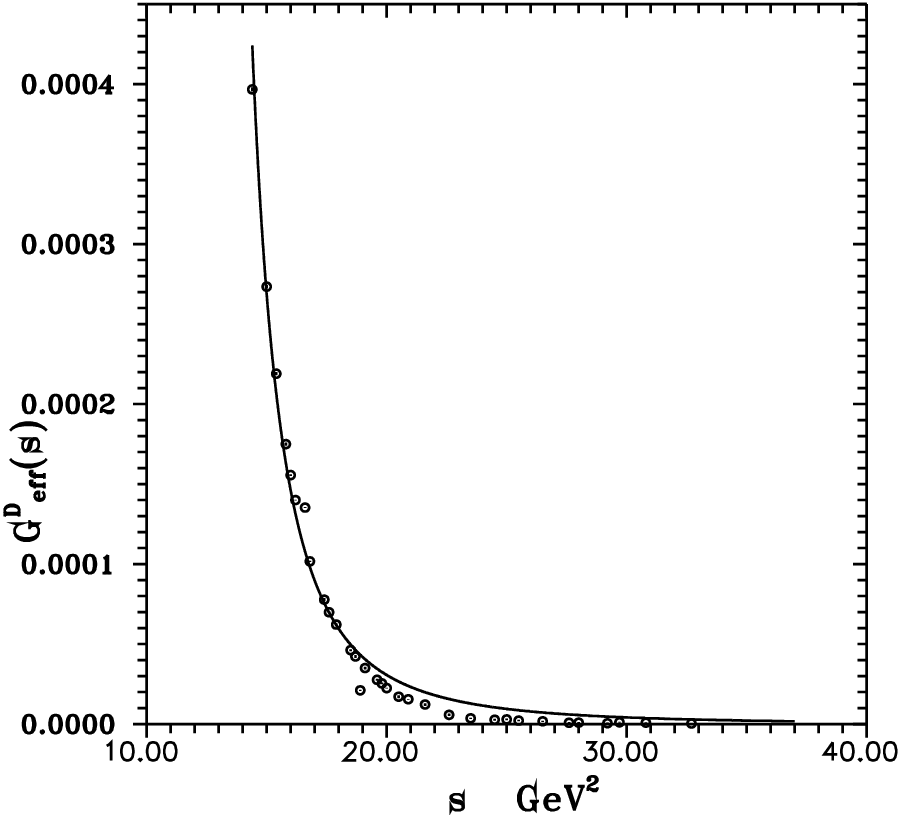}\\
  \caption{Artificial data on the "effective" deuteron EM FF scattered around the theoretically predicted curve by the formula (\ref{effFF}).}\label{FIG. 7:}
\end{figure}

\begin{figure}[bth]
  \includegraphics[width=.30\textwidth]{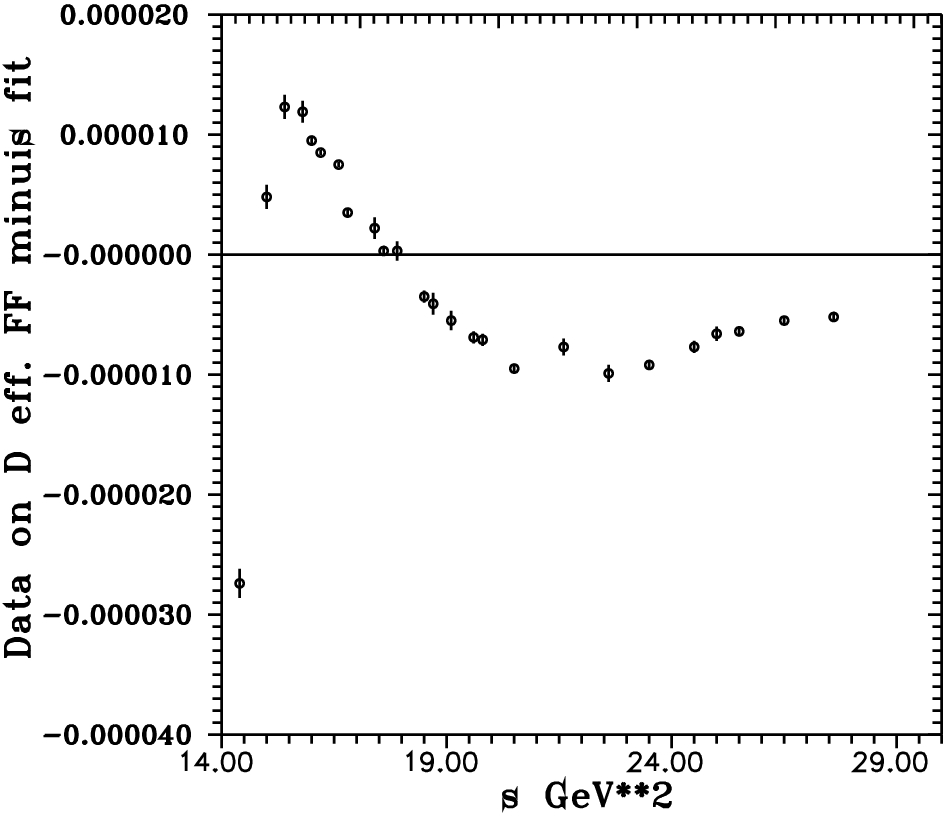}\\
  \caption{The deuteron damped oscillation regular structures obtained by a subtraction of the curve given in FIG. 7 from the numerical values of the artificial data on the deuteron "effective" EM FF with errors.}\label{FIG. 8:}
\end{figure}

   However, if the curve in FIG. 8, theoretically predicted by the U$\&$A model, is subtracted from the same artificial data on the deuteron "effective" EM FF with
errors, no expressive damped oscillation regular structures are revealed, as it is seen from Fig. 11.
\begin{figure}[bth]
  \includegraphics[width=.30\textwidth]{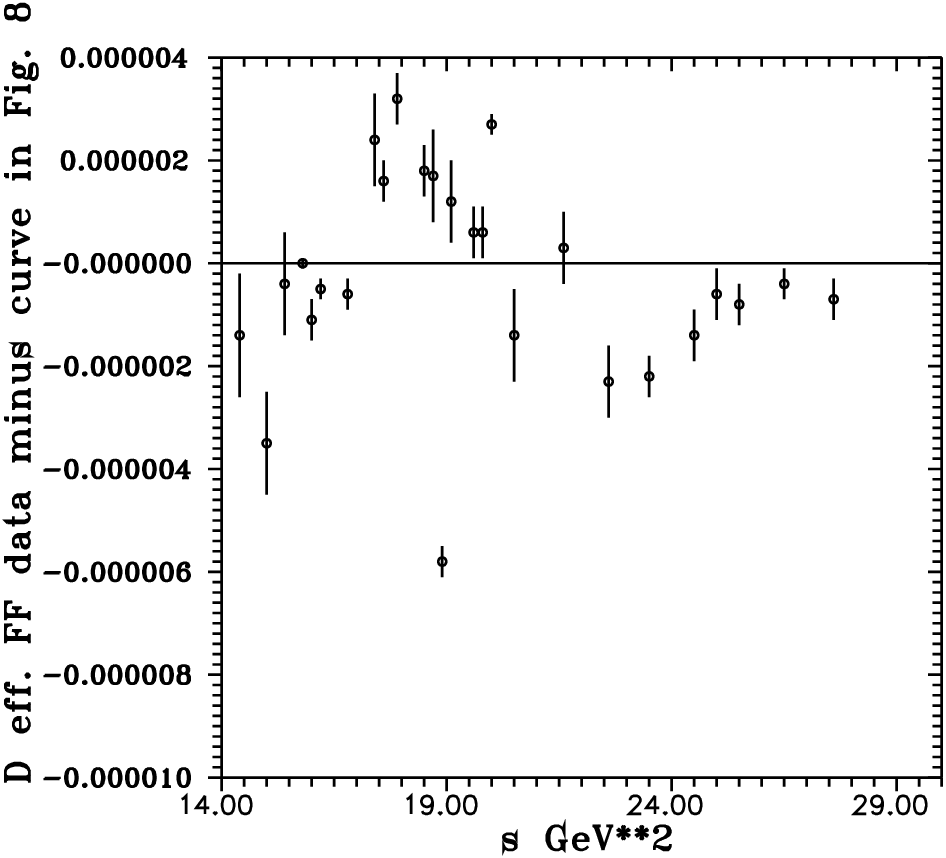}\\
  \caption{No expressive deuteron damped oscillation regular structures are seen by a subtraction of the curve in FIG. 7, predicted by the U$\&$A model of deuteron EM structure, from the numerical values of the artificial data on the deuteron "effective" EM FF with errors there.}\label{FIG. 11:}
\end{figure}

\section{CONCLUSIONS AND DISCUSSION}

   The RDOS from the proton "effective" form factor data in \cite{BTG} raised a wide interest to study the damped oscillatory regular structures also in the case of
other strongly interacting particles, for which solid data together with a physically well founded model for their accurate description exist.

   In this work the problem of the existence of the RDOS of the deuteron effective EM FF has been investigated using the same procedure as the one used in the case of
the proton in \cite{BTG}, despite of the fact that there is no experimental value on its $\sigma_{tot}(e^+e^- \to D \bar D)(s)$ up to now.

   When the "effective" data on the deuteron EM structure are described by the three parametric function of \cite{TGR},
the regular damped oscillatory structures appear. However, if for a description of the same data more physically founded U$\&$A model of the EM structure of hadrons is applied, no expressive regular damped oscillatory structures appear.

   So, the results of all our investigations \ref{BDD}-\ref{DDHL}, including also this paper, indicate that there is no objective existence of the damped oscillation
regular structures from the "effective" hadron EM FF data and their appearance is most probably due to application of the three parametric formula \cite{TGR}, without any physical background, for a description of data on the "effective" hadron EM FFs with the adequate accuracy. On the other hand, our U\&A model approach demonstrates its capabilities to be applicable for a perfect description of a wide range of hadronic processes without appearance of the damped oscillation regular structures.

\vspace{1cm}

The authors would like to thank Erk Bartos and Andrej Liptaj for valuable discussions and comments.

The second author acknowledges the support of the Slovak Grant Agency for Sciences VEGA, grant No.2/0084/25.

\end{document}